\documentclass[longtitle]{elsarticle}

\usepackage{epstopdf}
\usepackage{url}
\usepackage{rotating}

\newcommand{\pks}{PKS~2155-304}

\begin{document}

\begin{frontmatter}

\title{Search for Lorentz Invariance breaking with a likelihood fit of the \pks{} flare data taken on MJD~53944}

\author[ad1]{HESS Collaboration, A.~Abramowski}
\author[ad2]{F.~Acero}
\author[ad3,ad4,ad5]{F.~Aharonian}
\author[ad6,ad5]{A.G.~Akhperjanian}
\author[ad7]{G.~Anton}
\author[ad8,ad9]{A.~Barnacka}
\author[ad10]{U.~Barres~de~Almeida\fnref{fn1}}
\author[ad11]{A.R.~Bazer-Bachi}
\author[ad12,ad13]{Y.~Becherini}
\author[ad14]{J.~Becker}
% ...

\address[ad1]{
Universit\"at Hamburg, Institut f\"ur Experimentalphysik, Luruper Chaussee 149, D 22761 Hamburg, Germany
}

\address[ad2]{
Laboratoire de Physique Th\'eorique et Astroparticules, Universit\'e Montpellier 2, CNRS/IN2P3, CC 70, Place Eug\`ene Bataillon, F-34095 Montpellier Cedex 5, France
}

\address[ad3]{
Max-Planck-Institut f\"ur Kernphysik, P.O. Box 103980, D 69029 Heidelberg, Germany
}

\address[ad4]{
Dublin Institute for Advanced Studies, 31 Fitzwilliam Place, Dublin 2, Ireland
}

\address[ad5]{
National Academy of Sciences of the Republic of Armenia, Yerevan, Armenia 
}

\address[ad6]{
Yerevan Physics Institute, 2 Alikhanian Brothers St., 375036 Yerevan, Armenia
}

\address[ad7]{
Universit\"at Erlangen-N\"urnberg, Physikalisches Institut, Erwin-Rommel-Str. 1, D 91058 Erlangen, Germany
}

\address[ad8]{
Nicolaus Copernicus Astronomical Center, ul. Bartycka 18, 00-716 Warsaw, Poland
}

\address[ad9]{
CEA Saclay, DSM/IRFU, F-91191 Gif-Sur-Yvette Cedex, France
}

\address[ad10]{
University of Durham, Department of Physics, South Road, Durham DH1 3LE, UK
}

\address[ad11]{
Centre d'Etude Spatiale des Rayonnements, CNRS/UPS, 9 av. du Colonel Roche, BP 4346, F-31029 Toulouse Cedex 4, France
}

\address[ad12]{
Astroparticule et Cosmologie (APC), Universit\'{e} Paris 7 Denis Diderot, CNRS/IN2P3, CEA, Observatoire de Paris, 10 rue Alice Domon et L\'{e}onie Duquet, F-75205 Paris Cedex 13, France
}

\address[ad13]{
Laboratoire Leprince-Ringuet, Ecole Polytechnique, CNRS/IN2P3, F-91128 Palaiseau, France
}

\address[ad14]{
Institut f\"ur Theoretische Physik, Lehrstuhl IV: Weltraum und Astrophysik, Ruhr-Universit\"at Bochum, D 44780 Bochum, Germany
}

% ...

\cortext[cor]{
Corresponding author
}

\fntext[fn1]{%\dag
Supported by CAPES Foundation, Ministry of Education of Brazil
}

\fntext[fn2]{%\ddag
Present address: Kavli/SLAC, 2575 Sand Hill Road, Menlo Park, CA 94025, USA
}

\begin{abstract}
Several models of Quantum Gravity predict Lorentz Symmetry breaking at energy scales approaching the Planck scale ($\sim10^{19}$ GeV). With present photon data from the observations of distant astrophysical sources, it is possible to constrain the Lorentz Symmetry breaking linear term in the standard photon dispersion relations. Gamma-ray Bursts (GRB) and flaring Active Galactic Nuclei (AGN) are complementary to each other for this purpose, since they are observed at different distances in different energy ranges and with different levels of variability. Following a previous publication of the High Energy Stereoscopic System (H.E.S.S.) collaboration \cite{hessqg}, a more sensitive event-by-event method consisting of a likelihood fit is applied to \pks{} flare data of MJD 53944 (July 28, 2006) as used in the previous publication. The previous limit on the linear term is improved by a factor of $\sim$3 up to $\mathrm{M}^{l}_\mathrm{QG} > 2.1\times10^{18}$ GeV and is currently the best result obtained with blazars. The sensitivity to the quadratic term is lower and provides a limit of M$^{q}_\mathrm{QG} >$ 6.4$\times10^{10}$ GeV, which is the best value obtained so far with an AGN and similar to the best limits obtained with GRB.
\end{abstract}

\begin{keyword}
active galaxies \sep \pks \sep H.E.S.S. \sep quantum gravity \sep Lorentz invariance breaking
\end{keyword}

\end{frontmatter}

\section{Introduction}

\subsection{Lorentz Invariance and Quantum Gravity}

%%%%%%%%%% TABLE 1 %%%%%%%%%%%
%%%%%%%%%%%%%%%%%%%%%%%%%%%%%%

A development of new space-borne and ground-based experiments in the last decade, covering a large domain of gamma-ray astronomy, opened a new window for fundamental physics tests. In particular, the confrontation of the results from astrophysical sources with challenging predictions of various models in the frame of Quantum Gravity theories became of general interest \cite{amelino08}.  
The theory of Quantum Gravity (see \textit{e.g.} \cite{smolin01}), which is still incomplete, would give a unified picture based on both Quantum Mechanics and General Relativity, thus leading to a common description of the four fundamental forces.

Quantum Gravity effects in the framework of String Theory \cite{ellis99}, where the gravitation is considered as a gauge interaction, result from a graviton-like exchange in a classical space-time. In most String Theory models involving large extra-dimensions, these effects would take place at the Planck scale, thus not leading to ``spontaneous'' Lorentz Symmetry breaking, as may happen in models with ``foamy'' structure \cite{amelino98,ellis99} of quantum space-time. In this second class of models, photons propagate in a vacuum which may exhibit a non-negligible refractive index due to its foamy structure on a characteristic scale approaching the Planck length or equivalently the Planck energy $\mathrm{E}_\mathrm{P} = 1.22\times10^{19}$~GeV. This implies a group velocity of light changing as a function of energy of the photons, in analogy to the dispersion effects in any theoretical description of plasma media.

On the other hand, in models based on General Relativity with Loop Quantum Gravity \cite{alfaro02,gambini99} which postulate discrete (cellular) space-time in the Planckian regime, the fluctuations would introduce perturbations to the propagation of photons. The light wave going through the discrete space matrix would feel an induced perturbation which increases with decreasing wavelength or equivalently with increasing energy of photons. In consequence, photons with different wavelengths propagate with different velocity. 

As a result, one may expect a spontaneous violation of Lorentz Symmetry at high energies to be the generic signature of the Quantum Gravity.

In four dimensions, the Quantum Gravity scale is presumed to be close to the Planck mass and the standard photon dispersion relations up-to second order corrections in energy can be written as:
\begin{equation}
\label{eq:disp}
c^2 p^2 = E^2 \left( 1 \pm \xi(E/M) \pm \zeta(E/M)^2 \pm ... \right),
\end{equation}
where $\xi$ and $\zeta$ are positive parameters.

This implies an energy-dependent speed-of-light with
\begin{equation}
\label{eq:2}
v = c \left(1 - \xi(E/M)\right)
\end{equation}
considering only the linear term in the expansion. Alternatively assuming the linear correction equal to zero, the quadratic term leads to:
\begin{equation}
\label{eq:3}
v = c \left(1 - \zeta(E/M)^2\right).
\end{equation}

High-energy photons could propagate either slower (the sub-luminal case) or faster (the super-luminal case) than low-energy photons. In this study, a causality conserving scheme was adopted following to the theoretical arguments of the non-critical string models, developed in \cite{sarkar02}. The sign minus in equations \ref{eq:2} and \ref{eq:3} is required to avoid Cherenkov radiation in vacuum \cite{amelino97}. As results of this analysis and as in case of most of other quoted results in Table~\ref{tab:res}, only limits on Quantum Gravity scale for the sub-luminal photons will be given.

As suggested by Amelino-Camelia \cite{amelino98}, the tiny effects due to Quantum Gravity can add up to measurable time delays for photons from cosmological sources. The energy dispersion is best observed in sources that show fast flux variability, that are at cosmological distances and are observed over a wide energy range. This is the case of Gamma Ray Bursts (GRB) and Very High Energy (VHE) flares of Active Galactic Nuclei (AGN). Both types of sources are the preferred targets of these Òtime-of-flightÓ studies, which provide the least model-dependent tests of the Lorentz Symmetry. The case of pulsed emission by Pulsars has also been considered, and provided valuable results \cite{kaaret}.

It should be underlined that studying both GRB and AGN is of fundamental interest. GRB are observed with satellites at very large distances (up to $z\sim8$) but with very limited statistics above a few tens of GeV. On the contrary, AGN flares can be detected by ground based instruments with large statistics up to a few tens of TeV. Due to the absorption of the high energy photons by Extragalactic Background Light (EBL), TeV observations are limited to sources with low redshifts. Thus, GRB and AGN are complementary since they allow to test different energy and redshift ranges.

It should be noted that time-of-flight measurements are subject to a bias related to a potential dispersion introduced from the intrinsic source effects, which could cancel-out or enhance the dispersion due to modifications to the speed of light.

When considering sources at cosmological distances, the analysis of time lags as a function of redshift requires a correction due to the expansion of the universe \cite{jacob}, which depends on the cosmological model. Following the analysis of the BATSE data and recently of the \textit{HETE-2} and \textit{Swift} GRB data \cite{ellis03, ellis06, ellis08a, bolmont} and within a framework of the Standard Cosmological Model \cite{bahcall} with a flat expanding universe and a cosmological constant, the difference in the arrival time of two photons emitted at the same time is given by 
\begin{eqnarray}
\label{eq:linear}
\frac{{\rm\Delta} t}{{\rm\Delta} E} \approx \frac{\xi}{\mathrm{E}_{\rm P} \mathrm{H}_\mathrm{0}} \int_0^z
   \frac{(1+z')\,dz'}{\sqrt{\Omega_m(1+z')^3 + \Omega_{\Lambda}}}\end{eqnarray}
for a linear effect, and by
\begin{eqnarray}
\label{eq:quad}
\frac{{\rm\Delta} t}{{\rm\Delta} E^2} \approx \frac{3 \zeta}{2 \mathrm{E}^2_{\rm P} \mathrm{H}_\mathrm{0}} \int_0^z
   \frac{(1+z')^2\,dz'}{\sqrt{\Omega_m(1+z')^3 + \Omega_{\Lambda}}}\end{eqnarray}
for a quadratic effect. In the present study the cosmological parameters were set to $\Omega_m = 0.3$, $\Omega_{\Lambda} = 0.7$ and $\mathrm{H}_\mathrm{0} = 2.3\times10^{-18}$~s$^{-1}$. $\Delta E$ and $\Delta E^2$ represent linear and quadratic energy ranges respectively.

\subsection{Present results}

\begin{sidewaystable}
\caption{A selection of limits obtained with various instruments and methods for both GRB and AGN\label{tab:res}}
\begin{tabular}{llllll}
\hline
Source(s) & Experiment & Method & Results (95\% CL limits) & Reference & Note\\
\hline
GRB 021206     & \textit{RHESSI}                 & Fit + mean arrival time in a spike      & M$^{l}_\mathrm{QG} >$ 1.8$\times10^{17}$ GeV & \cite{boggs}  & $^a$,$^b$ \\
GRB 080916C    & \textit{Fermi} GBM + LAT        & associating a 13 GeV photon with the    & M$^{l}_\mathrm{QG} >$ 1.3$\times10^{18}$ GeV & \cite{fermi1} &   \\
               &                                 & trigger time                            & M$^{q}_\mathrm{QG} >$ 0.8$\times10^{10}$ GeV &               &   \\
GRB 090510     & \textit{Fermi} GBM + LAT        & associating a 31 GeV photon with the    & M$^{l}_\mathrm{QG} >$ 1.5$\times10^{19}$ GeV & \cite{fermi2} & $^c$ \\
               &                                 & start of any observed emission, DisCan & M$^{q}_\mathrm{QG} >$ 3.0$\times10^{10}$ GeV &               & $^c$ \\
9 GRBs         & BATSE + OSSE                    & wavelets   & M$^{l}_\mathrm{QG} >$ 0.7$\times10^{16}$ GeV & \cite{ellis03} & $^a$ \\
               &                                 &            & M$^{q}_\mathrm{QG} >$ 2.9$\times10^{6}$ GeV  &   &\\
15 GRBs        & \textit{HETE-2}                 & wavelets   & M$^{l}_\mathrm{QG} >$ 0.4$\times10^{16}$ GeV & \cite{bolmont} & $^d$ \\
17 GRBs        & \textit{INTEGRAL}               & likelihood & M$^{l}_\mathrm{QG} >$ 3.2$\times10^{11}$ GeV & \cite{lamon} & $^e$ \\
35 GRBs        & BATSE + \textit{HETE-2}         & wavelets   & M$^{l}_\mathrm{QG} >$ 1.4$\times10^{16}$ GeV & \cite{ellis06,ellis08a} & $^f$,$^g$ \\
               &  + \textit{Swift}               &            &                                              &  &\\
Mrk 421        & Whipple                         & likelihood & M$^{l}_\mathrm{QG} >$ 0.4$\times10^{17}$ GeV & \cite{biller} & $^a$,$^h$ \\
Mrk 501        & MAGIC                           & ECF        & M$^{l}_\mathrm{QG} >$ 0.2$\times10^{18}$ GeV & \cite{albert} &\\
               &                                 &            & M$^{q}_\mathrm{QG} >$ 0.3$\times10^{11}$ GeV &    &\\
               &                                 & likelihood & M$^{l}_\mathrm{QG} >$ 0.3$\times10^{18}$ GeV & \cite{martinez} &\\
               &                                 &            & M$^{q}_\mathrm{QG} >$ 5.7$\times10^{10}$ GeV &    &\\
PKS 2155-304   & H.E.S.S.                        & MCCF       & M$^{l}_\mathrm{QG} >$ 7.2$\times10^{17}$ GeV & \cite{hessqg} &\\
               &                                 &            & M$^{q}_\mathrm{QG} >$ 1.4$\times10^{9}$ GeV  &  &\\
               &                                 & wavelets   & M$^{l}_\mathrm{QG} >$ 5.2$\times10^{17}$ GeV &  &\\
               &                                 & likelihood & M$^{l}_\mathrm{QG} >$ 2.1$\times10^{18}$ GeV &  & $^i$\\
               &                                 &            & M$^{q}_\mathrm{QG} >$ 6.4$\times10^{10}$ GeV &  &\\
\hline
\end{tabular}
$^a${Limit obtained not taking into account the factor $(1+z)$ in the integral of equations \protect{\ref{eq:linear}} and \protect{\ref{eq:quad}}.}
$^b${The pseudo-redshift estimator \cite{pelan} was used. This estimator can be wrong by a factor of 2.}
$^c${Only the most conservative limits are given here. The best limits given in \cite{fermi2} which can still be considered as secure are M$^{l}_\mathrm{QG} >$ 10 $\mathrm{E}_\mathrm{P}$ and M$^{q}_\mathrm{QG} >$ 8.8$\times10^{10}$ GeV.}
$^d${Photon tagged data was used.}
$^e${The pseudo-redshift estimator \cite{pelan} was used for 6 GRB out of 11.}
$^f${For HETE-2, fixed energy bands were used.}
$^g${The limits of Ellis et al. \cite{ellis06} were updated in \cite{ellis08a} taking into account the factor $(1+z)$ in the integral of equation \protect{\ref{eq:linear}}. Only the limit obtained for a linear correction is given.}
$^h${A likelihood procedure was used, but not on an event-by-event basis.}
$^i${This work.}
\end{sidewaystable}

The most important results obtained so far with transient astrophysical sources are given in Table~\ref{tab:res}. The most robust limits to date have been obtained by Ellis et al. \cite{ellis06,ellis08a} and Bolmont et al. \cite{bolmont}, with the use of several GRB at different redshifts, resulting in limits of the order of M$^{l}_\mathrm{QG} > 10^{16}$~GeV. In case of a significant detection, using several sources at different redshifts allows in principle to take into account the possible time-lags originating from source effects. 

\textit{Fermi} is now in operation and has provided the best limits so far \cite{fermi1, fermi2}. However, for the moment the results are obtained with no study of the variation of the time-lag with redshift and with limited statistics in the GeV range. The recent studies with flares of Mrk~501 \cite{albert, martinez} and the former result of H.E.S.S. with \pks{} \cite{hessqg} provide results setting a lower limit on the linear Quantum Gravity scale at a few $10^{17}$ GeV for each source.

Table \ref{tab:res} also shows the different analysis techniques used to measure the energy-depend time-lags. The analysis methods can be divided in two categories, analyses applied on binned data on the one hand and unbinned analyses on the other hand.

Binned analyses are applied on binned light curves and consist of finding the position of local extrema in two different energy bands and deducing the value of the lag. A simple fit \cite{boggs} or a more sophisticated technique such as Continuous Wavelet Transform (CWT) can be used to localize extrema. CWT was used for GRB \cite{ellis03, ellis06, ellis08a, bolmont} and \pks{} \cite{hessqg}. The Modified Cross Correlation Function (MCCF) was also used for \pks{} to determine directly the lag from two light curves in two different energy bands \cite{hessqg}.

Unbinned analyses have also been used for both GRB and AGN. Likelihood fits require a parameterization of the light curves and this parameterization can be taken from a binned \cite{martinez} or unbinned fit \cite{lamon} to the light curve. Another method, the Energy Cost Function (ECF), was recently used by Albert et al. \cite{albert}. In this method one introduces a time lag $\tau$ for each photon and adds up their energies if they fall into a given time interval. The maximum of ECF($\tau$) gives the value of the measured lag. The recently developed method called DisCan \cite{scargle} was used for the analysis of Fermi data \cite{fermi2}. As discussed in \cite{scargle}, this method applies successfully to the analysis of very sharp peaks as can be found in GRB light curves.

It is important to point out here that the use of at least two methods is recommended since different methods can probe different aspects of the light curve. The present analysis completes previously published studies \cite{hessqg} where two different methods were employed.

As no significant time-lag was detected so far, careful error calibration studies using Monte Carlo simulations are mandatory for the extraction of the limits, as will be further discussed in the following.

%%%%%%%%%%%% TABLE 2 %%%%%%%%%%%%%%
%%%%%%%%%%%%%%%%%%%%%%%%%%%%%%%%%%%

\subsection{Overview of the paper}

In this paper, a new analysis of \pks{} data is presented, using a likelihood fit to determine both linear and quadratic corrections to the dispersion relations. In section~\ref{sec:method}, the method is described. The data in use as well as the selections applied are detailed in section~\ref{sec:data}. Then, in section~\ref{sec:cal}, the procedure used to calibrate the method and to evaluate systematics is presented. The results are given in section~\ref{sec:res} and finally discussed in section~\ref{sec:conc}.

\section{Method description}\label{sec:method}

To study the correlations between the arrival time and the energy of the photons, a method relying on a probability density function (p.d.f.) on an event-by-event basis is used. Following Martinez and Errando \cite{martinez}, the p.d.f. is defined by:
\begin{eqnarray}
\label{eq:pet}
P(t, E) = N \int_0^\infty A(E_{S})\Lambda(E_{S})G(E-E_{S},\sigma(E_{S})) F_{S}(t-\tau_l E_{S}) dE_{S},
\end{eqnarray}
where $\Lambda(E_{S})$ is the emission spectrum, $G(E-E_{S},\sigma(E_{S}))$ is the smearing function in energy (a usual point spread function for an energy resolution of 10\%), $A(E_{S})$ is the acceptance of the H.E.S.S. array, $F_{S}$ is a parameterization of the emission time distribution (or light curve) and $N$ is a normalization factor. the time shift was re-parameterized as $\Delta t = -\tau_l E$ and $\Delta t = -\tau_q E^2$ for linear and quadratic effects respectively, where $\tau_l$ and $\tau_q$ are expressed in \mbox{s TeV$^{-1}$} and \mbox{s TeV$^{-2}$}. The value of $N$ was computed in view of unbiased $\tau$-parameter estimation. It guarantees the correct normalization of the p.d.f. and takes into account its model dependence. Here, integration over measured domain of energy and time has been performed.

The function $F_S$ is a function of time at the source $t_S = t-\tau E_{S}$. The probability $P(t, E)$ depends only on the relative delay of photons for different energies. As the light curve at the emission time is not known, the best estimate of the time structure at the source can be taken from the time distribution of the  lowest energy photons. In the propagation models, the change of the QG scale induces basically a sliding effect in time of the light curve. This point has been checked to be valid by the Monte Carlo simulations described below.

An important assumption of this study should be underlined: the intrinsic light curves have been considered to have the same shape in different energy bands. As no lag is found, this can be partially justified by comparing the light curve at low energy (Fig. \ref{fig:2}) and in the full energy range of the instrument where the statistics is five times higher \cite{hesspks}. In both cases the light curve shows a five peak structure with maxima located roughly at the same times.

The likelihood function is given by the product of probability density functions over all photons:
\begin{equation}
\label{eq:l}
L = \prod_{i}P_{i}(t, E).
\end{equation}
The maximum of $L$ provides the time-lag parameter $\tau_l$ and $\tau_q$ for linear and quadratic models respectively.

The likelihood method is in principle very sensitive and allows a lag to be measured even if the number of selected photons is limited, which is not the case for wavelet or MCCF analyses. As a consequence of this high statistical sensitivity, it would be possible to probe a light curve locally, introducing a selection in time. The main aspect of the method to be taken into account is a need for a very precise parameterization of the photon distributions.

%%%%%%%%%%%%% TABLE 3 %%%%%%%%%%%%%%%%
%%%%%%%%%%%%%%%%%%%%%%%%%%%%%%%%%%%%%%

\begin{table}
\caption{Selections applied to the data\label{tab:cuts}}
\begin{tabular}{lr}
\hline
Selection & Number of selected events\\
\hline
Total sample                                                       &   8148 (100 \%)    \\
(1) = Time in 0--4000 s                                            &   7713  (95 \%)    \\
(2) = (1) and $\theta^2 < 0.005$ deg.$^2$                          &   5974  (73 \%)    \\
(2) and $E_{m}$ in 0.25--4.0 TeV                                   &   3526  (43 \%)    \\
(2) and $E_{m}$ in 0.25--0.28 TeV                                  &    561   (7 \%)    \\
\hline
\end{tabular}
\end{table}

\section{Data sample used in the analysis}\label{sec:data}

\pks{} is a high-peaked BL Lac object located at redshift $z = 0.116$. On July 28, 2006 (MJD 53944), an extreme flare of this source was observed by H.E.S.S. \cite{hesspks}. The full data sample was obtained using a combination of two analysis \mbox{methods} \cite{denaurois_pal}: the standard Hillas analysis \cite{hillas} and the Model analysis \cite{denaurois_icrc}. This technique greatly improves the signal to background ratio and results in 8148 on-source events recorded in \mbox{$\sim$85 minutes} (three data-taking runs) with an energy threshold of $\sim$120~GeV.
The signal-to-background ratio in the flare data exceeds 300. The effect of the background contribution will be investigated in the next section. The average zenith angle of selected events is $\theta_z \approx 11^\circ$.

Table~\ref{tab:cuts} summarizes the different cuts applied to the data. In this study, only the first 4000~s of data (7713 events) are considered, where the flux and variability are the highest. To further improve the signal-to-background ratio, a cut $\theta^2 < 0.005$~deg.$^2$ is applied. This cut reduces the number of photons to 5974 events but allows a very good fit of the light curve and of the energy spectrum. A total of 3526 events pass the cut on $\theta^2$ in the first 4000~s of data in the range 0.25--4.0~TeV. All events in the data sample fulfilling the selection were used in the analysis.

\begin{figure}
\includegraphics[width=0.9\textwidth]{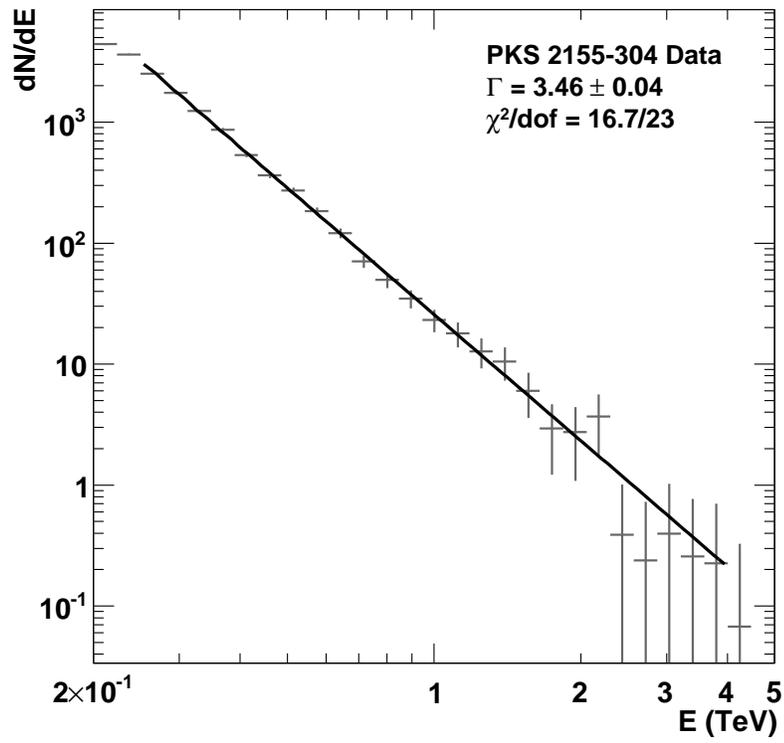}
\caption{\label{fig:1}Spectrum of PKS~2155-304 flare of MJD 53944, taking into account the cut on the off-axis angle mentioned in the text. Points are fitted with a power law $E^{-\Gamma}$.}
\end{figure}

As shown in Figure \ref{fig:1}, the reconstructed spectrum is compatible with a power law of index $\Gamma = 3.46$ in the range of 0.25--4.0 TeV (with $\chi^2/\mathrm{dof} = 16.7/23$). This result is compatible with the high energy part of the broken power law spectrum obtained for the same flare by Aharonian et al. \cite{hesspks} with a different event selection and energy reconstruction method, and also with the spectrum obtained for the whole 2005--2007 period \cite{pks0507}. The dependence of the spectral index as a function of time has been studied extensively in \cite{hesspks} and no significant variation was found during the flare.

\begin{figure}
\includegraphics[width=0.9\textwidth]{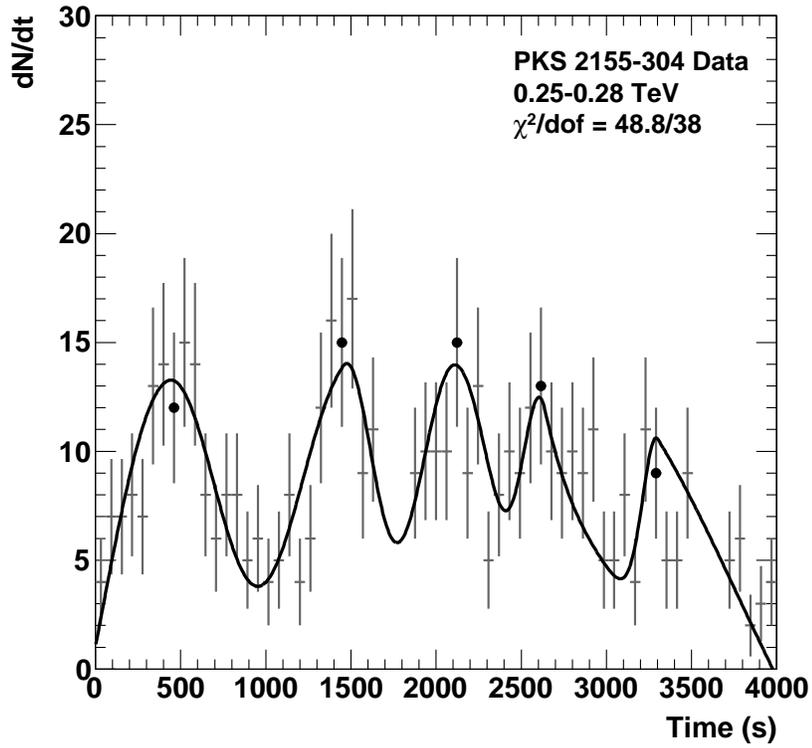}
\caption{\label{fig:2}Light curve of PKS~2155-304 flare of MJD 53944 in the range 0.25--0.28 TeV, with a bin width of 61~s and taking into account the cut on $\theta^2$ described in the text. Black points show the positions of the extrema as determined by the peak-finding procedure of Morhac et al.  \cite{morhac}. The light curve is fitted with a function $F_5(t) + B$ where $B$ is a constant and $F_5$ is defined by equation~\ref{eq:paramlc}. The zero of the time axis corresponds to MJD 53944.024.\vspace{1em}}
\end{figure}

Figure \ref{fig:2} shows the light curve in the range 0.25--0.28 TeV with a binning of 61~s. The shape of this light curve will be used later as a template for the time-lag determination (\S\ref{sec:res}). To estimate the main parameters of the light curve (widths and asymmetry of the pulses, distance between consecutive spikes), a parameterization was performed with a sum of asymmetric Gaussian functions. For this, the number of spikes was first determined using a peak-finding procedure based on a Markov chain algorithm \cite{morhac}. Five pulses were found and their positions were used as initial parameters for the light curve fit.

%%%%%%%%%%%%%% FIG 1 %%%%%%%%%%%%%%%%
%%%%%%%%%%%%%%%%%%%%%%%%%%%%%%%%%%%%%

%%%%%%%%%%%%%%%%%%%%% FIG 2 %%%%%%%%%%%%%%%%%%%%%%
%%%%%%%%%%%%%%%%%%%%%%%%%%%%%%%%%%%%%%%%%%%%%%%%%%

The fit function $F$ is a sum of asymmetric Gaussian curves, defined as follows:
\begin{equation}
\label{eq:paramlc}
F_{n}(t) = \sum_{i=1}^{n} f_i(t, A_i, \mu_i, \alpha_i, \beta_i),
\end{equation}
where $n$ is the number of spikes, and where
\begin{equation}
\label{eq:asymgauss}
f(t, A, \mu, \alpha, \beta) =
\left \{
\begin{array}{l l l l}
    y & = & A\ e^{-\frac{(t - \mu)^2}{2\alpha^2}},& \mathrm{if\ t} < \mu\\
    y & = & A\ e^{-\frac{(t - \mu)^2}{2\beta^2}},& \mathrm{if\ t} \geq \mu. \\
\end{array}
\right.
\end{equation}
In this definition, $A$ and $\mu$ are the normalization factor and the position of the pulse and $\alpha$ and $\beta$ are the left-hand and right-hand widths of the pulse respectively. The asymmetry of the curve can be quantified by the parameter $\nu = \beta/\alpha$. When $\nu = 1$, $f$ becomes a Gaussian function with a width $\alpha = \beta = \sigma_0$.

\begin{table}
\caption{Parameters of the function $F_5(t) + B$ fitted on PKS~2155-304 light curve in the range 0.25--0.28 TeV\label{tab:parampks}}
\begin{tabular}{lrr}
\hline
Parameter & Value & Error\\
\hline
$A_1$          & 27.1   &  1.4   \\
$\mu_1$        & 430.4  &  35.8  \\
$\alpha_1$     & 399.9  &  25.4  \\
$\beta_1$      & 352.9  &  75.7  \\
$A_2$          & 26.0   &  0.8   \\
$\mu_2$        & 1456.1 &  16.0  \\
$\alpha_2$     & 342.9  &  92.8  \\
$\beta_2$      & 193.3  &  44.9  \\
$A_3$          & 28.0   &  0.7  \\
$\mu_3$        & 2113.1 &  19.2  \\
$\alpha_3$     & 276.1  &  46.2  \\
$\beta_3$      & 296.8  &  29.3  \\
$A_4$          & 20.0   &  0.7   \\
$\mu_4$        & 2639.7 &  77.0  \\
$\alpha_4$     & 130.3  &  34.3  \\
$\beta_4$      & 899.8  &  23.7  \\
$A_5$          & 9.4   &  1.3   \\
$\mu_5$        & 3302.3 &  19.9  \\
$\alpha_5$     & 87.1  &  36.5  \\
$\beta_5$      & 999.9  &  80.4 \\
$B$            & $-14.1$  &  0.8   \\
\hline
\end{tabular}
\end{table}

The parameterization was carried out with a function $F_5$ plus a constant term $B$. Values of the parameters are given in Table~\ref{tab:parampks}, leading to $\chi^2/\mathrm{dof} = 48.8/38$. This fit gives the following information about the light curve:
\begin{itemize}
\item width of the pulses in the range 250--600~s, 
\item time difference between consecutive spikes in the range 500--1000~s, and
\item ratio $\nu$ in the range 0.6--7.
\end{itemize}

These intervals were investigated for the calibration procedure described in the next section.

\section{Calibration of the method}\label{sec:cal}

A calibration procedure was developed using a toy Monte-Carlo program in order to reflect as closely as possible the real data analysis and to provide the best strategy to minimize the systematic errors on the estimated time lag. 

As discussed in the previous section, the light curve of \pks{} is well fitted by a superposition of five asymmetric Gaussian spikes. The energy spectrum follows a power law distribution. These parameters have been used as initial conditions for the simulation. The zenith angle was considered to be constant at $\theta_z = 10^\circ$.

A simulation procedure comprises the following steps:
\begin{itemize}
\item the template $F_S$ is chosen to be either a Gaussian function or derived from the measured light curve in the range 0.25--0.28 TeV;
\item photons are generated from a parameterizations $F_S$ and $\Lambda$ in the range 0.25--4 TeV;
\item for each photon, a lag $\Delta t$ is introduced depending on the energy;
\item a weight is associated to each photon taking into account the acceptance of the detector during data-taking in 2006;
\item the energy is smeared taking into account the H.E.S.S. energy resolution ($\Delta E/E \sim$10\%);
\item when a simple Gaussian light curve is generated, histograms obtained for the light curve and the spectrum are fitted. As a result, the functions used for the fits reproduce well those used for the generation. This point was checked to estimate the effects of the statistical smearing, which were found to be negligible;
\item finally, the likelihood is computed and the minimum of $-2\Delta\ln(L)$ is searched for, first using a Gaussian template parameterization for the detailed studies or the \pks{} template. As shown later, the distribution of the minima has always been found to be compatible with a Gaussian distribution (with a mean $\overline{\tau}$ and a width $\sigma_\tau$).
\end{itemize}

For each simulation run, these steps are repeated 500 times with $\sim$2000 photons after selections in each realization. The injected lag ($\tau_0$) range was chosen between $-100$ and 100~s\,TeV$^{-1}$ (respectively s\,TeV$^{-2}$) with steps of 10~s\,TeV$^{-1}$ (s\,TeV$^{-2}$).

\begin{figure}
\includegraphics[width=0.5\textwidth]{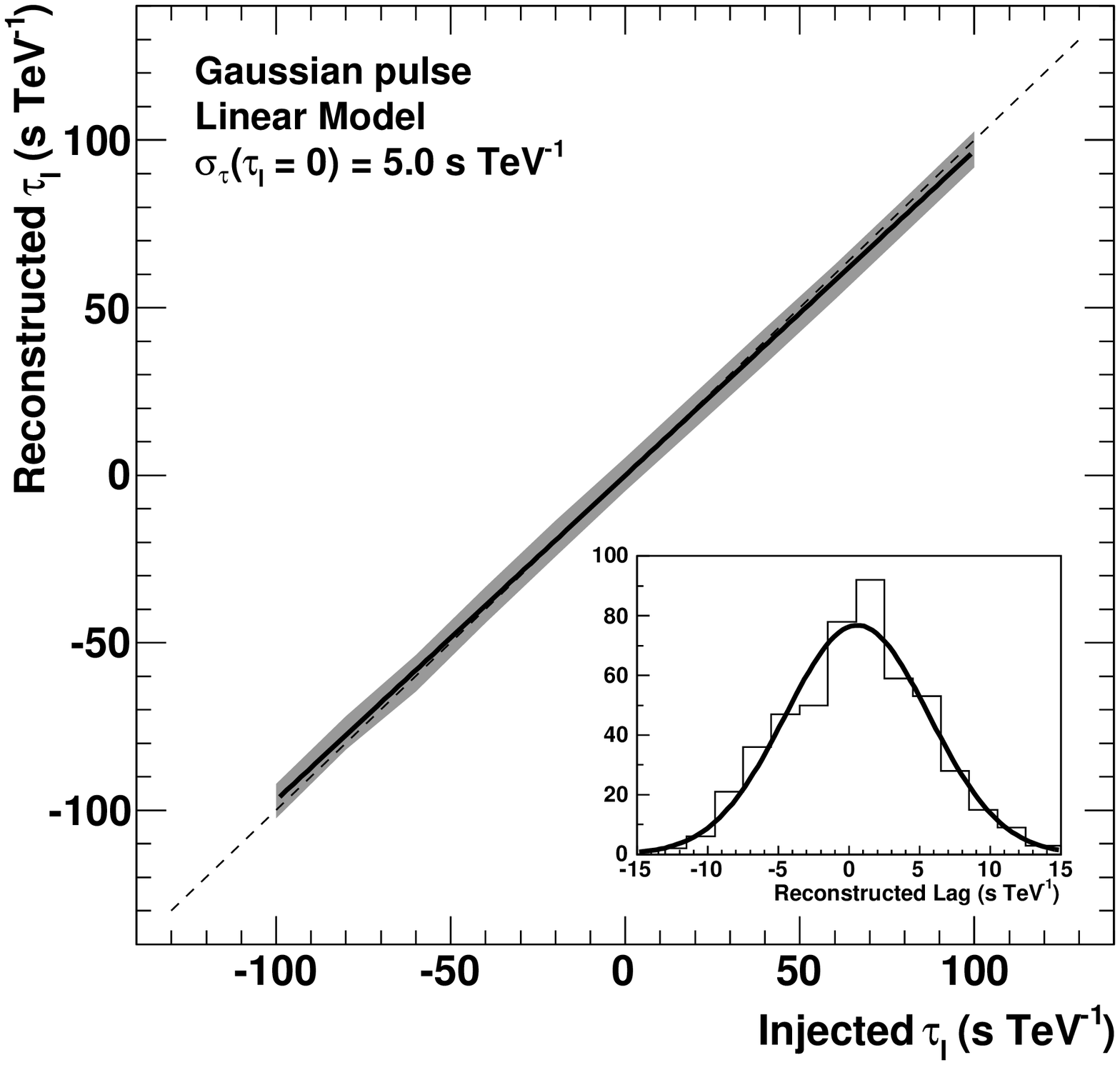}
\includegraphics[width=0.5\textwidth]{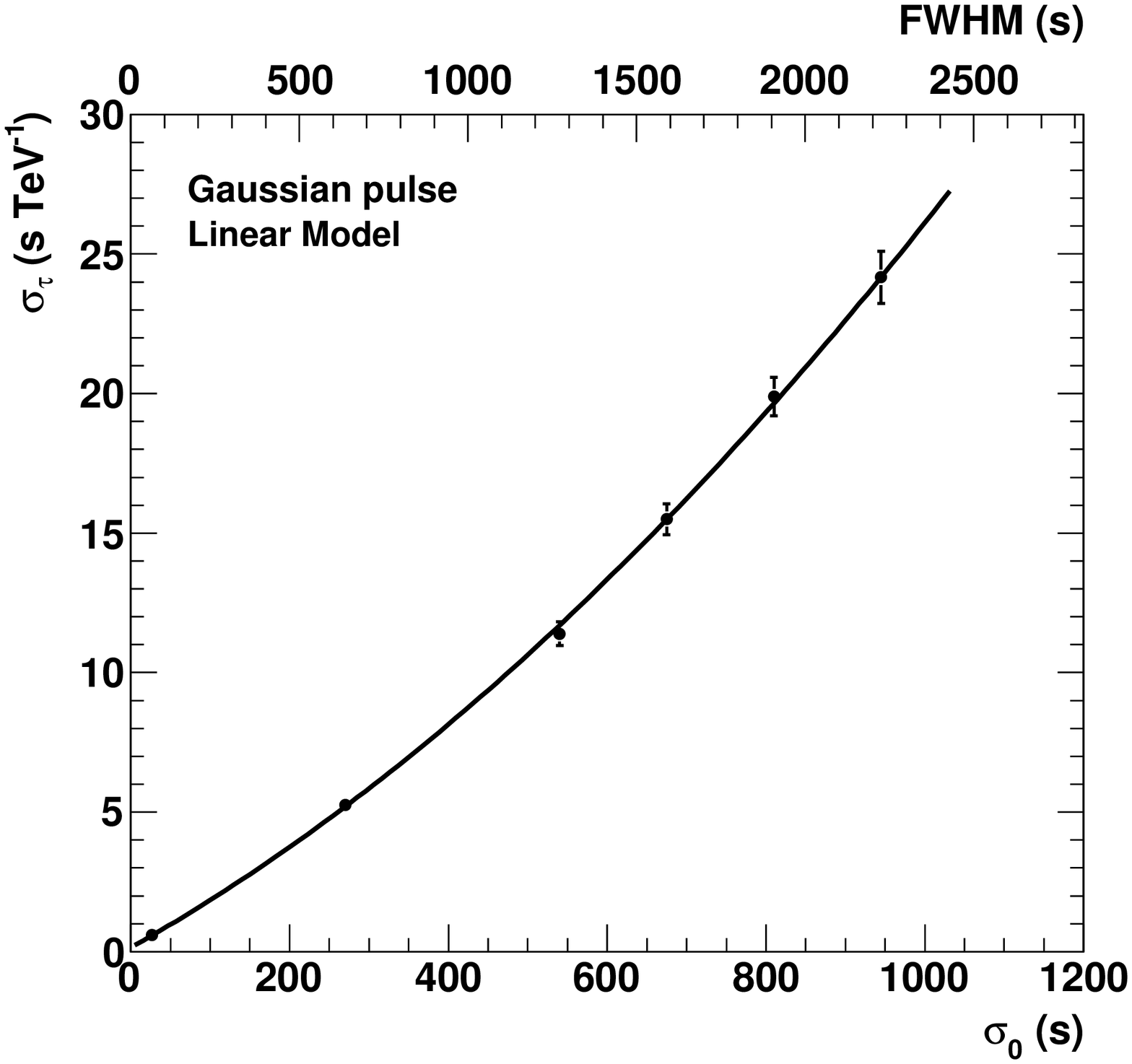}
\caption{\label{fig:3}Left: reconstructed lag as a function of the injected lag for a Gaussian light curve in the modelling mode (68\% CL range). The plot in the bottom right corner shows the distribution of the minima for the 500 realizations with a generated lag of 0 s\,TeV$^{-1}$. The dashed line shows the linear response function. Right: parameter $\sigma_\tau$ as a function of the width of the Gaussian pulse. The axis at the top gives the full width at half maximum (FWHM) of the pulse.}
\end{figure}

The results of the simulations can be summarized studying the variation of the reconstructed lag $\overline{\tau}$ as a function of the injected lag. Figure~\ref{fig:3} (left) shows such a plot for simulation settings which will be described later (Gaussian spike).

The shaded area corresponds to a 68\% CL range of the reconstructed lag. A fit within $\pm\sigma_\tau$ yields a slope parameter value close to unity. %Due to a large number of realizations, the error on the slope parameter is negligible.

A small systematic shift of the slope is observed; however, the bias on the reconstructed time lag is much smaller than the statistical uncertainty $\sigma_\tau$ for time lags between \mbox{$-$50~s\,TeV$^{-1}$} and \mbox{50~s\,TeV$^{-1}$}.

In the following sub-sections, different features of the error calibration procedure are detailed. First, the influence of the pulse shape (width and asymmetry) and the superposition of pulses is investigated. Then, the dependence on the spectral index is evaluated. In the end, the effect of background events is studied. The simulation of light curves with five pulses provide the value of the calibrated statistical error which is used in \S\ref{sec:res} to compute the limits on $\mathrm{E}_\mathrm{QG}$.

\begin{figure}
\includegraphics[width=0.5\textwidth]{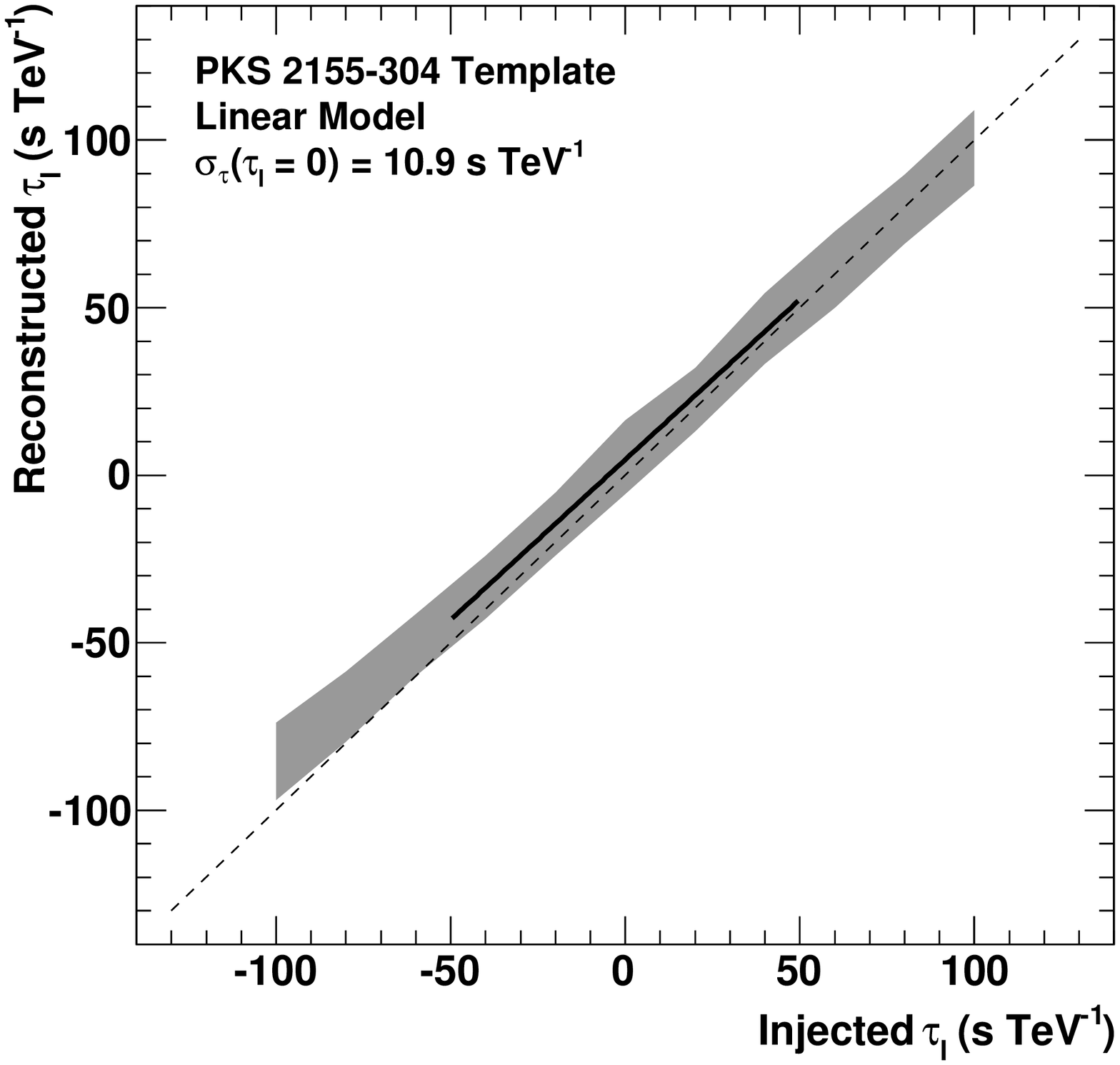}
\includegraphics[width=0.5\textwidth]{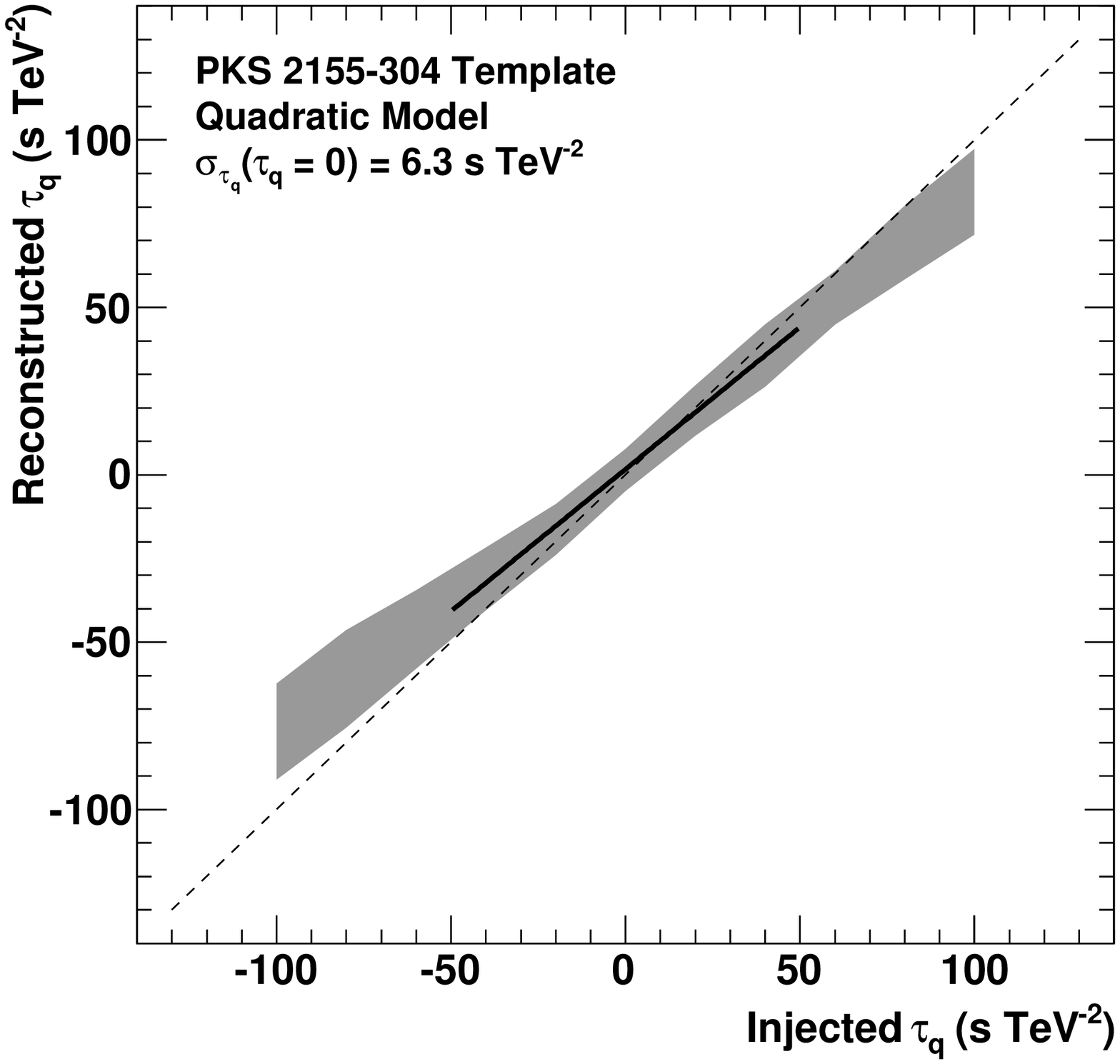}
\caption{\label{fig:4}Reconstructed lag (68\% CL range) as a function of the injected lag for a light curve $F_5(t) + B$ obtained from real data and for a linear (left) and a quadratic (right) effect. The dashed line shows the linear response function.}
\end{figure}

%%%%%%%%%%%%%%%%%%%%%% FIG 4 %%%%%%%%%%%%%%%%%%%%%%%
%%%%%%%%%%%%%%%%%%%%%%%%%%%%%%%%%%%%%%%%%%%%%%%%%%%%

%%%%%%%%%%%%%%%%%% FIG 3 %%%%%%%%%%%%%%%%%%%%
%%%%%%%%%%%%%%%%%%%%%%%%%%%%%%%%%%%%%%%%%%%%%

\subsection{Pulse shape and superposition of pulses}\label{subsec:shape}

%%%%%%%%%%%%%%%%%%%%%% FIG 5 %%%%%%%%%%%%%%%%%%%%%%%
%%%%%%%%%%%%%%%%%%%%%%%%%%%%%%%%%%%%%%%%%%%%%%%%%%%%

Figure~\ref{fig:3} (left) shows the reconstructed lag as a function of the injected lag for a Gaussian pulse of a mean width $\sigma_0$ of 300~s for a linear model. The calibration slope is $0.97\pm0.02$, showing the likelihood fit provides efficient means for the reconstruction of a lag. No deviation from linearity is observed in this case. The same figure shows an example of the distribution of the minima of $-2\Delta\ln(L)$ fitted with a Gaussian curve, for an injected lag of \mbox{0 s\,TeV$^{-1}$}. This distribution is fitted by a Gaussian curve with $\overline{\tau} = 0.6\pm0.2$~s\,TeV$^{-1}$ and \mbox{$\sigma_\tau = 5.1\pm0.2$ s\,TeV$^{-1}$}.

For a given value of the pulse width $\sigma_0$, the spread $\sigma_\tau$ is very stable. On the other hand, as shown in Figure~\ref{fig:3} (right), $\sigma_\tau$ increases rapidly with the width of the pulse. For a single pulse width of 600~s, $\sigma_\tau$ rises to $\sim13$ s\,TeV$^{-1}$.

When assuming asymmetry in the pulse shape, the calibration slope remains very stable and close to unity. The maximal asymmetry factor observed for \mbox{\pks{}} is $\nu = 7$, leading to \mbox{$\sigma_\tau \approx 10$ s\,TeV$^{-1}$} for $\sigma_0$ of 300~s. 

The superposition of spikes requires detailed studies concerning the evolution of the calibration slope and the error in case of complex light curves such as the one measured with \pks{} flare. In order to check the effect of spike superposition, light curves were generated from the sum of several asymmetric Gaussian curves. Various configurations were tested and the resulting calibration curves were compared to those obtained when injecting the real \pks{} light curve. 

The plots of Figure~\ref{fig:4} were obtained with light curves generated from real data in the range 0.25--0.28 TeV for the linear and quadratic modelling. These simulations were used to investigate the effect of  superimposing two to five pulses with realistic parameters leading to similar calibrated error values. 

%The slope and values of $\sigma_\tau$ are somewhat lower in the quadratic case. The lower values of $\sigma_\tau$ do not hold for the larger values of the injected time-lags, for which a deviation from the linearity of the calibration factors is observed. 

In the quadratic case, the slope is  somewhat smaller than 1. The spread $\sigma_\tau$ is smaller than in the linear case close to $\tau_0 = 0$ but increases at larger values of $|\tau_0|$ where the calibration curve becomes non-linear.

In the linear case, a systematic shift of the calibration values of the order of $1/2\sigma_\tau$ is observed. This effect will be included in the overall calculation of the systematic errors (see Table~\ref{tab:systematic}).

The error values obtained from Figure \ref{fig:4} were used in the following to derive the \mbox{limits} for linear and quadratic corrections. The error $\sigma_\tau$ was found to depend mainly on individual pulse widths rather than on the addition of the widths of the five pulses, leading to 10.9 s TeV$^{-1}$ for the linear and 6.3 s TeV$^{-2}$ for the quadratic case respectively. Anticipating the results from data, these values were taken in the vicinity of the injected time-lag $\tau_0 = 0$.

In addition, other functional forms were used when fitting template light curve: Norris parameterization as also used in \cite{hesspks} and Landau function. No detectable effect was found when deriving the results. Moreover, the effect of the different functional forms was evaluated with MC simulations.

\begin{figure}
\includegraphics[width=0.9\textwidth]{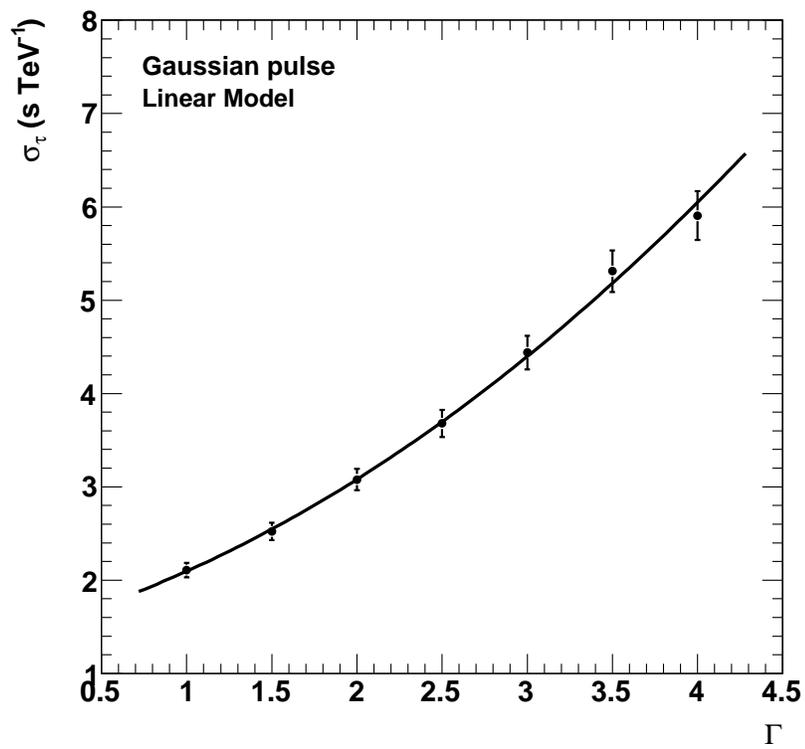}
\caption{\label{fig:5}Parameter $\sigma_\tau$ as a function of the spectral index.}
\end{figure}

\subsection{Dependence on spectral index}

Varying the spectral index of the generated light curves from 1 to 4.5 resulted in no significant change in the calibration slope (below 5\%) for a single Gaussian pulse. The systematic error due to the uncertainty on the spectral index determination can then be considered as negligible. In contrast, the error value is a strong function of the spectral index as also shown in Figure~\ref{fig:5}. %The value of $\sigma_\tau$ increases with the spectral index. 
A harder spectrum gives a higher precision, due to the fact that the photons are spread more uniformly between high and low energies. This enhances the potential of studies with closer AGN such as Mrk 421 and Mrk 501.

%%%%%%%%%%%%%% TABLE 4 %%%%%%%%%%%%%%%%
%%%%%%%%%%%%%%%%%%%%%%%%%%%%%%%%%%%%%%%

\begin{sidewaystable}
\caption{Systematic uncertainties of the event-by-event likelihood\label{tab:systematic}}
\begin{tabular}{llll}
\hline
             & Estimated error & Change in estimated $\tau_l$ (s TeV$^{-1}$) & Change in estimated $\tau_q$ (s TeV$^{-2}$)\\
\hline
Selection cuts          &                           & $<$ 5  & $<$ 5\\
Background contribution & 1\%                       & $<$ 1  & $<$ 1\\
Acceptance factors      & 2\%                       & $<$ 1  & $<$ 1\\
Energy resolution       & 1\%                       & $<$ 1  & $<$ 1\\
Energy calibration      & 10\%                      & $<$ 2  & $<$ 2\\
Spectral index          & 1\%                       & $<$ 1  & $<$ 1\\
Calibration systematics (constant, shift) & 10\%    & $<$ 5  & $<$ 1\\

$F_S(t)$ parameterization  &                        & $\approx$7 & $\approx$3\\
Total                      &                        & $<$ 10.3 & $<$ 6.6 \\
\hline
\end{tabular}
\end{sidewaystable}

\subsection{Effect of background contribution}\label{subsec:background}

Even though the signal to background ratio is very high for the studied data sample, it is possible that a small contribution of mis-identified photons introduces a small bias in the measurement of the time-lag. To study the impact of this effect, light curves were simulated with a fraction of particles with no energy-dependant time-lag as is expected for charged cosmic-rays. A realistic power law spectrum with an index of 2.7 was used to generate these particles. When 1\% of the particles are considered as protons, no significant change in either slope or parameter $\sigma_\tau$ is observed.

\subsection{Summary of calibration studies}

In this section, the most important features of the calibrations with the likelihood method are summarized.

As the emission light curve at the source is not known, the function $F_S$ was derived from real data in the range 0.25--0.28 TeV. In this case, it is shown that the likelihood fit can reconstruct the injected lag extremely well. In the future, if a satisfying emission model is proposed, the likelihood approach can be very useful to test propagation and emission models as well.

The error on the measured lag ($\sigma_\tau$) depends strongly on the width of individual pulses.This result is very important and allows competitive results to be obtained with the \pks{} light curve. On the other hand, the asymmetry of the pulses and the superposition of several pulses has been shown to have less important effect on the results. The parameter $\sigma_\tau$ also depends on the spectral index, which is another source of uncertainty. However, for \mbox{\pks} flare in 2006, the spectral index can be determined with a very good precision. 

The systematic uncertainties in the method employed were quantified by varying selections and cuts and treating the light curve shape parameterization in different ways. The various contributions are detailed in Table~\ref{tab:systematic}. The main uncertainty was found to be related to event selections and to the light curve parameterizations. 

In order to get a precise estimation of the systematic uncertainty related to the $F_S$ parameterization, the covariance matrix of the fit presented in Figure~\ref{fig:2} was studied. The values of the fit parameters were varied following to the gaussian distributions with parameters derived from the covariance matrix after its diagonalization. The values in Table~\ref{tab:systematic} were obtained with this procedure for the linear and the quadratic case separately. 

The overall systematic error is estimated to be \mbox{$< 10.3$ s\,TeV$^{-1}$} and \mbox{$< 6.6$ s\,TeV$^{-2}$} for the linear and quadratic effects respectively. These values will be combined with the statistical calibrated error given in \S\ref{subsec:shape} when determining the Quantum Gravity scales.

%%%%%%%%%%%%%%%%%%%% FIG 6 %%%%%%%%%%%%%%%%%%%%%%
%%%%%%%%%%%%%%%%%%%%%%%%%%%%%%%%%%%%%%%%%%%%%%%%%

\section{Results with PKS~2155-304 flare of MJD~53944}\label{sec:res}

\begin{figure}
\includegraphics[width=0.5\textwidth]{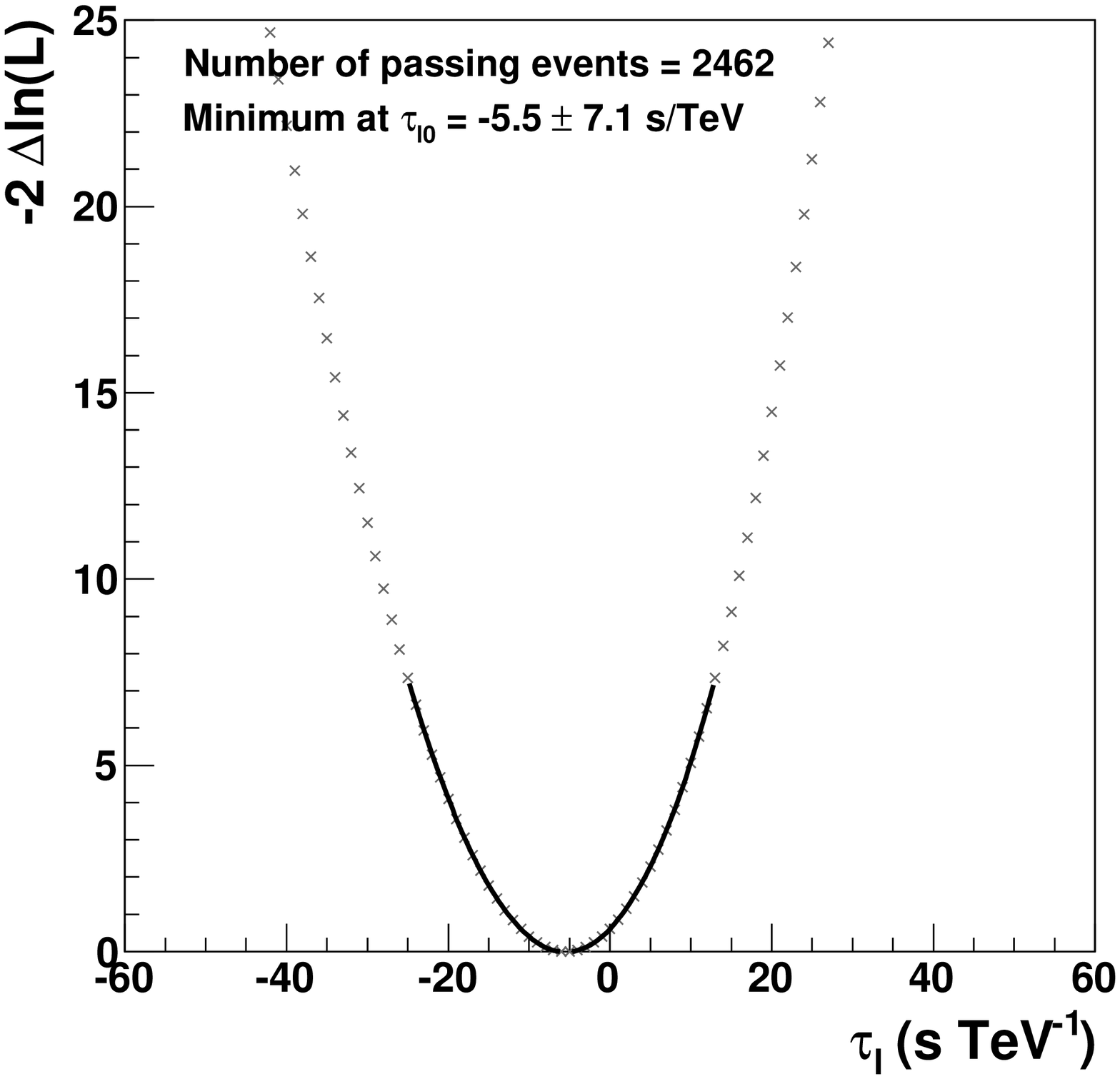}
\includegraphics[width=0.5\textwidth]{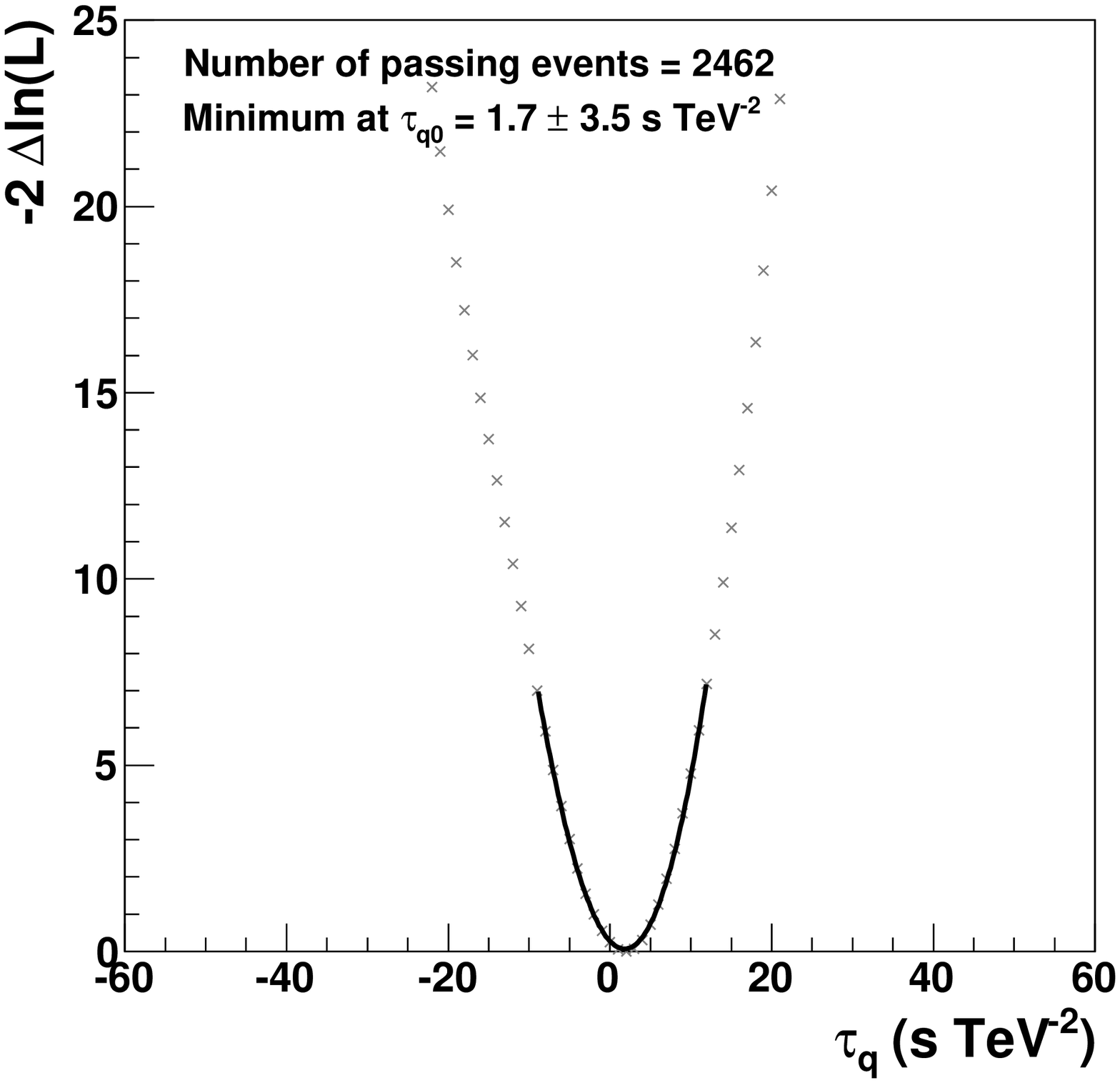}
\caption{\label{fig:6}$-2\Delta\ln(L)$ as a function of $\tau$ when the measured light curve is fitted in 0.25--0.28 TeV and the likelihood is computed in 0.3--4.0 TeV for a linear (left) and quadratic (right) correction. Points are fitted with a third-degree polynomial with a minimum at $\tau_{l0} = -5.5\pm{7.1}$ s\,TeV$^{-1}$ and $\tau_{q0} = 1.7\pm{3.5}$ s\,TeV$^{-2}$. The errors on these values are obtained requesting $-2\Delta\ln(L) = 1$.}
\end{figure}

The final result was obtained with a parameterization of the light curve in the range 0.25-0.28 TeV ($\overline{E_1} = 0.26$~TeV, see light curve in Figure~\ref{fig:2}) and a likelihood computed in 0.3-4.0 TeV (\mbox{$\overline{E_2} = 0.49$~TeV}). The choice of these ranges was dictated by the balance found between lever-arm in energy and the statistics used in the likelihood fit. These choices were studied and optimized with the simulations described above.

Figure~\ref{fig:6} shows the $-2\Delta\ln(L)$ curve as a function of the injected $\tau$ for the linear and quadratic cases, leading to the value for the minimum of 
\begin{displaymath}
\tau'_{0l} = -5.5 \pm 10.9_\mathrm{(stat)} \pm 10.3_\mathrm{(sys)}\ \mathrm{s\,TeV}^{-1}
\end{displaymath}
for the linear and 
\begin{displaymath}
\tau'_{0q} = 1.7 \pm 6.3_\mathrm{(stat)} \pm 6.6_\mathrm{(sys)}\ \mathrm{s\,TeV}^{-2}
\end{displaymath}
for the quadratic effect.

These results imply that there is no significant time-lag measured. Consequently, the 95\% CL limits on the Quantum Gravity energy scales are derived:
\begin{displaymath}
\mathrm{M}^{l}_\mathrm{QG} > 2.1\times10^{18}\ \mathrm{GeV}\ (\xi < 5.7)
\end{displaymath}
and 
\begin{displaymath}
\mathrm{M}^{q}_\mathrm{QG} > 6.4\times10^{10}\ \mathrm{GeV}\ (\zeta < 3.6\times10^{16}).
\end{displaymath}

The more conservative 99\% CL limits are 
\begin{displaymath}
\mathrm{M}^{l}_\mathrm{QG\ 99\%} > 1.3\times10^{18}\ \mathrm{GeV}\ (\xi < 9.2)
\end{displaymath}
and 
\begin{displaymath}
\mathrm{M}^{q}_\mathrm{QG\ 99\%} > 5.3\times10^{10}\ \mathrm{GeV}\ (\zeta < 5.2\times10^{16}).
\end{displaymath}

These results were found to be very stable when considering different energy domains and time selections. The stability studies of the estimated systematic effects \mbox{allows} the quoted values to be considered as conservative.

The limit on the linear term, while lower by a factor of ten in comparison with the one obtained by \textit{Fermi} with GRB 090510 \cite{fermi2}, shows that with AGN too, the conclusion about the models predicting a linear effect with energy \cite{ellis99} tends to be confirmed. As far as the limit on the quadratic term is concerned, it is the best one obtained so far with AGN and GRB, even if it still remains far from the Planck scale value.

As stated in the introductory part, the presented paper focused on models proposed in \cite{ellis03,ellis06} where predictions of linear Lorentz Invariance Violation effects were expected and the causality condition was conserved. Moreover, it may be noticed that because of negligible negative values found both for the linear and quadratic terms, the super-luminal case limits will differ only slightly from those provided for the sub-luminal case.

\section{Summary and prospects}\label{sec:conc}

The limits obtained in this study are the most constraining ever obtained with an AGN. They are almost ten times higher than those obtained by Martinez and Errando \cite{martinez} with the same method for the Mrk 501 flare recorded by MAGIC, and consistent with their rough estimate for the PKS 2155-304 flare observed by H.E.S.S.. They are also a factor of $\sim$3 higher than the previous H.E.S.S. result \cite{hessqg}. The increase in sensitivity as compared to the MAGIC result is mainly due to excellent parameters of the data taken on July 28, 2006: low zenith angle ($\sim10^\circ$) which leads to a low energy threshold and high statistics with high variability. On the other hand, the steepness of the \pks{} energy spectrum was one of the penalizing factors in this study.

Given the complexity of the \pks{} light curve, a detailed calibration study of the method was necessary. For this purpose, a toy Monte-Carlo program was developed. The simulations allowed the influence of different parameters on the likelihood performance to be investigated, such as the shape of the light curve or spike superposition. In particular, a very interesting conclusion for a light curve with multiple spikes is that the error on the lag depends mainly on the width of the individual pulses. This feature of the likelihood fit allows a smaller error on the lag measurement to be obtained, yielding higher limits on M$_\mathrm{QG}$.

It should still be noticed that the superposition of different spikes leads to a decrease in sensitivity due to a limited number of ``useful events'' which contribute to the likelihood fit. Despite this loss in statistical power, the method allows robust and precise results on the fitted time-lag to be obtained.

Another important conclusion about the method is the possibility to obtain robust results even with low photon statistics. The use of different time selections showed that a large decrease in the number of selected events entering the fit had only a small influence on the error value of the time-lag measurement.

The main difficulty of time of flight studies with photons from AGN and GRB is that the origin of the measured lag is unknown. If the measured time-lag is not related to Quantum Gravity induced propagation effect, it can be due to an emission effect at the source. The measured lag can also be a superposition of source and propagation effects. As no lag was found in the studied sample, either the source effects compensate exactly the propagation effects (``conspiracy scenario''), or there are no source or propagation induced effects detected within present experimental sensitivity. In this study the second hypothesis was retained, which leads to a conclusion that any measured lag would be entirely due to Lorentz Invariance Violation effects.

In future, modelling of the emission light curve could bring new insight in the propagation and emission time studies. Population studies of AGN examining the effect of the distance on the lag measurement could allow this assumption to be avoided but would require that intrinsic effects are similar for different AGN. Better results on propagation effects will then come in parallel with a better understanding of the emission models.

More and more results on Lorentz symmetry breaking are published which give limits for the linear correction that are very close to the Planck energy scale. In particular, the latest results from \textit{Fermi} observations of distant GRB provide exclusion limits for a large class of linear models above the Planck energy \cite{fermi1,fermi2}. However, as pointed out by Ellis et al. \cite{ellis09}, `extraordinary claims require extraordinary evidence', so more studies will be needed in the future to give more robust results and to be able to definitively reject or validate proposed models.

The time-lag measurements with photons are far from being sensitive enough to constrain the quadratic term of the dispersion relations and do not exclude any model in this case. Multi-messenger and multi-wavelength campaigns will be necessary in future to constrain this quadratic term. Combining results obtained with GRB and AGN detected by Fermi and ground-based detectors would allow considering common scheme with photon energies ranging from the sub-GeV range to the multi-TeV range and from sources with higher redshift values. Future observations with neutrino telescopes of GRB and AGN flares would further extend the energy range to extreme, multi-hundred TeV domain. However, at present, the multi-messenger and multi-wavelength procedures are still to be developed.

Concerning the increase of the experimental potential in observations of the AGN, a key step towards a better understanding of the emission and the propagation effects will be achieved with a new generation of instruments, which will allow the observation of a larger number of AGN flares. HESS-II, with an energy threshold of $\sim$30 GeV and later the large Cherenkov array CTA\,\footnote{\url{http://www.cta-observatory.org/}} will increase dramatically the sensitivity to energy-dependent time-lags. A rough estimation obtained by extrapolating results from the \pks{} flare to the extended domain in energy from a few GeV to a hundred TeV, and assuming a gain in sensitivity of a factor of ten, would rise the limit on the quadratic term by two orders of magnitude. This type of results may be further confronted with present and future theoretical models. It may also be underlined here that this will open a new unexplored domain for Lorentz Invariance Violation searches.

\section{Acknowledgments}
The support of the Namibian authorities and of the University of Namibia
in facilitating the construction and operation of H.E.S.S. is gratefully
acknowledged, as is the support by the German Ministry for Education and
Research (BMBF), the Max Planck Society, the French Ministry for Research,
the CNRS-IN2P3 and the Astroparticle Interdisciplinary Programme of the
CNRS, the U.K. Science and Technology Facilities Council (STFC),
the IPNP of the Charles University, the Polish Ministry of Science and 
Higher Education, the South African Department of
Science and Technology and National Research Foundation, and by the
University of Namibia. We appreciate the excellent work of the technical
support staff in Berlin, Durham, Hamburg, Heidelberg, Palaiseau, Paris,
Saclay, and in Namibia in the construction and operation of the
equipment.

%\begin{figure*}

%\begin{figure*}

%\begin{figure*}

\end{document}